\documentclass[PRM,twocolumn,groupedaddress,nofootinbib,superscriptaddress]{revtex4-1}
\usepackage{graphicx}
\usepackage{dcolumn}
\usepackage{bm}
\usepackage{times}
\usepackage{epstopdf}
\usepackage[version=3]{mhchem} 
\usepackage{xcolor}

\usepackage{mathrsfs}
\usepackage{lipsum}
\usepackage{amsmath}
\usepackage{amsfonts}
\usepackage{array}
\usepackage{color}
\usepackage{ulem}

\newcommand{\degreesC}{\,$^{\circ}$C}

\begin{document}

\title{Double-layer Kagome Metals \ce{Pt3Tl2} and \ce{Pt3In2}}

\author{Michael A. McGuire}
\affiliation{Materials Science and Technology Division, Oak Ridge National Laboratory, Oak Ridge, Tennessee, U.S.A.}
\author{Eleanor M. Clements}
\affiliation{Materials Science and Technology Division, Oak Ridge National Laboratory, Oak Ridge, Tennessee, U.S.A.}
\author{Qiang Zhang}
\affiliation{Neutron Scattering Division, Oak Ridge National Laboratory, Oak Ridge, Tennessee, U.S.A.}
\author{Satoshi Okamoto}
\affiliation{Materials Science and Technology Division, Oak Ridge National Laboratory, Oak Ridge, Tennessee, U.S.A.}

\begin{abstract}
The connectivity and inherent frustration of the kagome lattice can produce interesting electronic structures and behaviors in compounds containing this structural motif.
Here we report the properties of Pt$_3$\textit{X}$_2$ ($X$\,=\,In and Tl) that adopt a double-layer kagome net structure related to that of the topologically nontrivial high temperature ferromagnet \ce{Fe3Sn2} and the density wave hosting compound \ce{V3Sb2}.
We examined the structural and physical properties of single crystal \ce{Pt3Tl2} and polycrystalline \ce{Pt3In2} using x-ray and neutron diffraction, magnetic susceptibility, heat capacity, and electrical transport measurements, along with density functional theory calculations of the electronic structure.
Our calculations show that Fermi levels lie in pseudogaps in the densities of states with several bands contributing to transport, and this is consistent with our Hall effect, magnetic susceptibility, and heat capacity measurements. While electronic dispersions characteristic of simple kagome nets with nearest-neighbor hopping are not clearly seen, likely due to the extended nature of the Pt $5d$ states, we do observe moderately large and non-saturating magnetoresistance values and quantum oscillations in the magnetoresistance and magnetization associated with the kagome nets of Pt.
\end{abstract}

\maketitle

\section{Introduction}

The kagome lattice, a 2D network of corner sharing triangles, has played an important role in materials physics for decades. For nearest neighbor antiferromagnetic interactions the lattice provides strong geometrical frustration, and kagome systems became important in the field of frustrated magnetism \cite{ramirez1994strongly}. This naturally led to keen interest in kagome lattice compounds as candidate quantum spin liquids, a key example being herbertsmithite, \ce{ZnCu3(OH)6Cl2} \cite{shores2005structurally, helton2007spin, han2012fractionalized, norman2016colloquium, mendels2016quantum}. Quite recently, renewed interest in the kagome lattice has been sparked by its relatively simple but physics-rich \textit{electronic} structure. The lattice symmetry and connectivity produce Dirac nodes, van Hove singularities, and flat bands. Notably, the flat bands are a result of frustrated electron hopping, reminiscent of the frustrated antiferromagnetic interactions the lattice hosts. Thus, kagome compounds provide model systems to explore the effects of these electronic structure features on a material's behavior when the Fermi level is located near them. Two clear motivations are the study of electronic topology and the effects of the large density of states associated with the flat band. Topological and Chern insulating behavior is expected under certain circumstances \cite{guo2009topological, sun2011nearly, okamoto2022topological}, and flat bands may provide a way of generating strong electron-electron interactions and correlated ground states \cite{balents2020superconductivity, sales2022chemical}.

Of particular relevance to the present study are the binary intermetallic compounds containing kagome layers like that shown in Figure \ref{fig:structure}c, typically made up of a transition metal \textit{M} (on the kagome net) and a main group metal \textit{X} \cite{sales2019electronic, kang2020dirac}. If these $M_3X$ layers are simply stacked the \ce{Mg3Cd} structure type is realized. This structure type is adopted by many compounds, including the high temperature ferromagnets \ce{Fe3Sn} and \ce{Fe3Ge} \cite{jannin1963magnetism, sales2014ferromagnetism, kanematsu1963magnetic, mcguire2018tuning} and chiral antiferromagnets \ce{Mn3Sn} and \ce{Mn3Ge} \cite{yang2017topological, liu2017anomalous, nayak2016large}. Alternating these kagome layers with honeycomb layers of the main group element (Figure \ref{fig:structure}d) results in the CoSn structure type in which the kagome nets are isolated from one another. Examples include FeGe, FeSn, CoSn, NiIn, RhPb, and PtTl, and these compounds have been studied in the context of flat bands, magnetism, and topology \cite{sales2019electronic, kang2020dirac, meier2019reorientation, meier2020flat, teng2022discovery, teng2023magnetism}. Materials with a third type of stacking have also received recent attention, those with two adjacent kagome layers separated by single honeycomb layers. This structure type is represented by two materials: the topologically nontrivial high temperature ferromagnet \ce{Fe3Sn2} \cite{fenner2009non, kida2011giant, ye2018massive} and the compound \ce{V3Sb2}, which undergoes a density wave transition near 160\,K \cite{wang2022density}. Recently several Ti-based compounds of this type were predicted theoretically \cite{yi2023topological}.

There is, in addition, the \ce{Pt3Tl2} structure type, a variant of the \ce{Fe3Sn2} structure with the same stacking sequence but a distinct stacking arrangement (Figures \ref{fig:structure}e and \ref{fig:structure}f). This structure is adopted by \ce{Pt3Tl2} and \ce{Pt3In2}. The structures of these compounds were reported more than fifty years ago \cite{bhan1968mischungen}, but no information about their electronic or magnetic properties were found in the literature. Here we examine the behavior of these two compounds using x-ray and neutron diffraction, density functional theory, and electrical transport, magnetization, and heat capacity measurements. We find that the Fermi level resides in a pseudo-gap in the density of states, and this theoretical result is in good agreement with our heat capacity and magnetization data. Multiple bands and carrier types appear to contribute to the conduction based on the calculated electronic structure and measured Hall effect. Due in part to the low density of states, we observe diamagnetic behavior, and we see no signature of phase transitions of any kind between room temperature and 2\,K. The anisotropic magnetoresistance in \ce{Pt3Tl2} at 2\,K, longitudinal and transverse, shows no sign of saturation up to 13.5\,T and reach values exceeding 200\%. The longitudinal magnetoresistance shows quantum oscillations (Shubnikov-de Haas oscillations), which are also seen in isothermal magnetization data (de Haas-van Alphen oscillations), indicating reasonably high mobility carriers in the materials.

\section{Materials and Methods}

Starting materials were Pt tubes and sheets (Refining Systems, 99.9\%), In shot (Alfa Aesar99.999\%) and Tl rods (Alfa Aesar 99.99\%). The Pt was cut into 1-2 mm-sized pieces before use to reduce time to react/dissolve during the syntheses. The In was used as received. The Tl was stored, cut, and weighted inside a He or Ar filled glove box to minimize oxidation. Note that Tl is toxic and should be handled with care to prevent exposure. All reactions were performed in alumina crucibles sealed inside fused silica ampoules under vacuum.

\begin{figure}
\begin{center}
\includegraphics[width=3.25in]{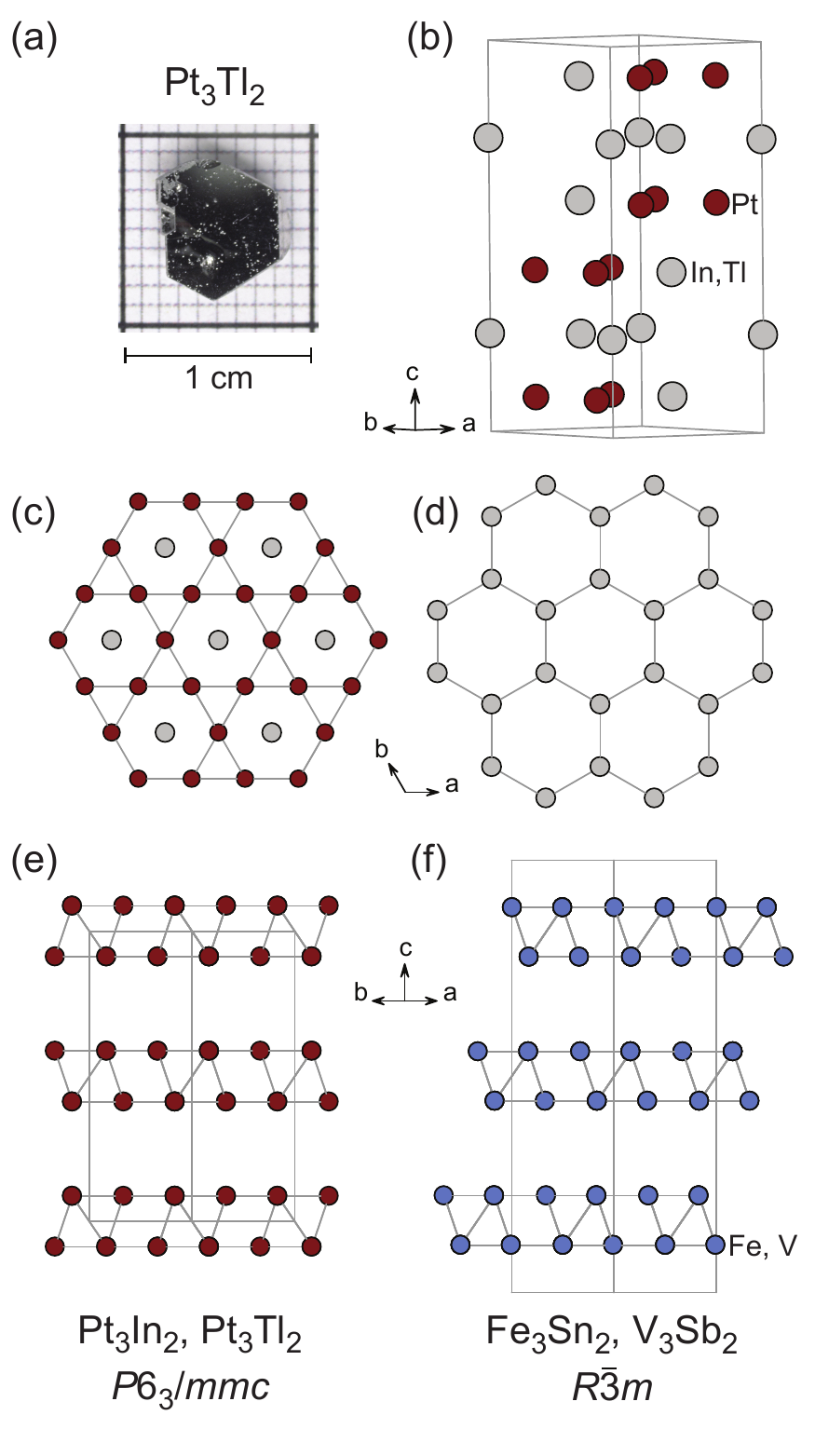}
\caption{\label{fig:structure}
Crystal structure of Pt$_3$\textit{X}$_2$ for \textit{X} = In and Tl. (a) A crystal of \ce{Pt3Tl2} grown for this study showing the hexagonal growth habit. (b) The hexagonal unit cell showing layers of composition Pt$_3X$ and $X2$. (c) A single Pt$_3X$ layer made up of Pt and $X3$ sites and containing a kagome net of Pt. (d) The honeycomb net made up of $X1$ and $X2$ sites that separates the double kagome slabs. (e) A view of the Pt$_3$\textit{X}$_2$ structure showing only the Pt atoms and highlighting the double kagome slabs. (f) For comparison, a view highlighting the double kagome slabs found in structures of \ce{Fe3Sn2} and \ce{V3Sb2} with only the transition metal sites shown.
}
\end{center}
\end{figure}

\ce{Pt3Tl2} decomposes peritectically at 900\degreesC. Single crystals of \ce{Pt3Tl2} were grown using the self-flux technique based on the published Pt-Tl binary phase diagram \cite{PtTlPD}. Pt and Tl were loaded into a Canfield crucible set \cite{Canfield2016} in a 45:55 molar ratio. At this ratio, the liquidus is near 880\degreesC. \ce{Pt3Tl2} is in equilibrium with the liquid down to 752\degreesC, at which point PtTl begins to precipitate. The reaction was heated over about 4 h to 950\degreesC, held at this temperature for 48 h to obtain a homogeneous melt, cooled to 800\degreesC\ at a rate of 2.3\degreesC/h, then inverted into a centrifuge to separate the \ce{Pt3Tl2} crystals from the liquid phase. Figure \ref{fig:structure}a shows a crystal grown in this way. Polycrystalline \ce{Pt3Tl2} was synthesized by first heating a stoichiometric mixture of the elements to 850\degreesC\ for several days, then 1080\degreesC\ for several hours. The sample was removed from the hot furnace into lab air to cool, then heated to 850\degreesC\ and held for five days. The sample was then ground, pressed into a pellet, and annealed at 850\degreesC\ for five days.

The Pt-In binary phase diagram is significantly more complicated than the Pt-Tl phase diagram \cite{PtInPD}. \ce{Pt3In2} is reported to be stable only below a peritectoid reaction at 550\degreesC. Thus, crystals are not attainable by self-flux techniques, and only polycrystalline \ce{Pt3In2} is studied here. This was made by reacting a stoichiometric mixture of the elements to 1050\degreesC\ for 3 days, then 525\degreesC\ for 6 days. Unreacted Pt was noted when grinding the product, so the sample was heated to 1100\degreesC\ for two days then quenched in water. The sample was then ground, pelletized, and annealed at 525\degreesC\ for one week.

Powder x-ray diffraction data were collected using a PANalytical X'Pert Pro MPD with monochromatic Cu $K_{\alpha1}$ radiation in the Bragg-Brentano geometry. The neutron diffraction experiment was conducted on the time-of-flight powder diffractometer, POWGEN, located at the Spallation Neutron Source of the Oak Ridge National Laboratory \cite{huq2011powgen, huq2019powgen}. The sample environment adopted was a POWGEN Automatic Changer. Approximately 1.1 g of powder sample was loaded into a 6 mm diameter vanadium can filled with helium gas.  Neutron bank with center wavelength of 1.5\,{\AA} was employed to collect the high-resolution neutron diffraction data covering a Q range of 0.48$-$12.9 {\AA}$^{-1}$. FullProf was used for Rietveld refinement of x-ray and neutron diffraction data \cite{Fullprof}. Energy dispersive spectroscopy (EDS) was performed using a Bruker Quantax 70 detector with a Hitachi TM3000 scanning electron microscope for semiquantitative chemical analysis.

Magnetization, electrical transport, and heat capacity data were measured using commercial cryostats from Quantum Design (PPMS, MPMS-3, Dyancool). For magnetization measurements, both Vibrating Sample Magnetometer and DC SQUID options from Quantum Design were used. For electrical transport (resistivity, Hall effect, and magnetoresistivity), Resistivity and AC-Transport options from Quantum Design were used. Electrical contacts to the samples were made using silver paste and platinum wires. All transport measurements were done with four contact points, and 2-point resistances between leads were confirmed to be $\sim 1 \Omega$ before measurements, and contacts were separated by distances on the order of 1\,mm. Heat capacity was measured using the relaxation method with samples mounted using Apiezon N-grease. Property measurements vs temperature were done on cooling. Isothermal magnetization measurements were done starting from the maximum positive fields. Magnetoresistance and Hall effect data were measured from maximum positive field to maximum negative field and then back to maximum positive field. The data were then appropriately symmetrized or antisymmetrized to remove offset voltages and voltages arising from misalignment of the contacts.

To gain insight into the electronic properties of Pt$_3$In$_2$ and Pt$_3$Tl$_2$, we carried out density functional theory (DFT) calculations
using  the Vienna Ab initio Simulation Package (VASP) \cite{Kresse1996,Kresse1999}, which uses
the projector augmented wave (PAW) approach \cite{Blochl1994} with the generalized gradient approximation in the parametrization of
Perdew, Burke, and Ernzerhof \cite{Perdew1996} for exchange correlation. Throughout the DFT calculations, the spin-orbit coupling was turned on.
To calculate the electronic ground state of these compounds,
we used the structural data obtained in this work (space group \# 194: $P6_3/mmc$),
a regular $12 \times 12 \times 6$ $\rm \bf k$-point mesh centered at the $\Gamma$ point,
an energy convergence criterion $10^{-6}$~eV, and an energy cutoff $500$ eV.
For Pt, we used a standard PAW potential (Pt in the VASP distribution),
while for In and Tl we used PAW potentials in which the $d$ semi core states are treated as valence states (In$_d$ and Tl$_d$ in the VASP distribution).
After the electronic ground state was obtained, we carried out non-selfconsistent calculations
to compute the density of states (DOS) using a denser $20 \times 20 \times 10$ $\rm \bf k$-point mesh with the smearing parameter $0.1$~eV
and the electronic band structure along high-symmetry lines using the optimized charge density.

\section{Results and Discussion}

X-ray diffraction confirmed that the samples adopt the previously reported structures \cite{bhan1968mischungen}. This hexagonal structure is illustrated in Figure \ref{fig:structure}b. It can be decomposed into layers in the ab-plane of composition Pt$_3X$ and of composition $X_2$. These layers are shown in Figures \ref{fig:structure}c and \ref{fig:structure}d, respectively. The Pt$_3X$ layer contains a kagome net of Pt sites with $X$ atoms at the centers of the hexagonal holes. The $X_2$ layer is a honeycomb net. In the full structure, two Pt$_3X$ layers are stacked together, forming a double kagome net of Pt, and these double layers are separated from one another by a single $X_2$ layer.

\begin{figure}
\begin{center}
\includegraphics[width=3.25in]{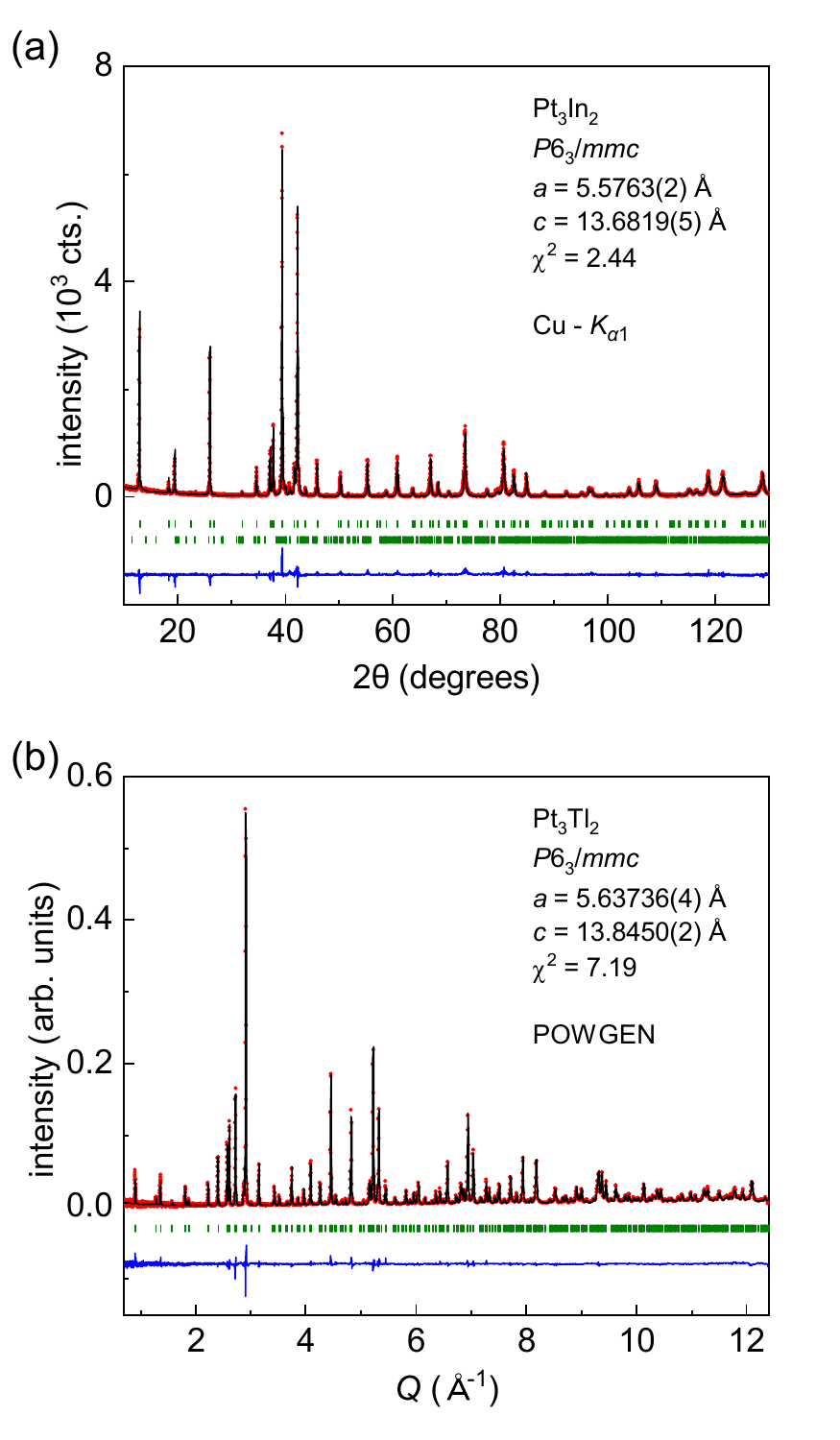}
\caption{\label{fig:diffraction}
Rietveld refinement of powder diffraction data from Pt$_3$\textit{X}$_2$ showing measured data points (red), fitted curves (black), and tick marks (green) locating Bragg reflections, along with difference curves (blue) at the bottom of each panel. (a) X-ray diffraction results for \ce{Pt3In2}. The upper ticks correspond to the main phase, and the lower ticks correspond to a \ce{Pt13In9}, an secondary phase present at the level of approximately 3 weight percent. (b) Time-of-flight neutron diffraction results for \ce{Pt3Tl2}. Both data sets were collected at room temperature.
}
\end{center}
\end{figure}

\begin{table}
\caption{\label{table} Crystal structure refinement results for Pt$_3$\textit{X}$_2$ using data collected at room temperature. The space group is $P6_3/mmc$. Pt is at $(x, 2x, z)$. $X1$ is at $(1/3, 2/3, z)$. $X2$ is at $1/3, 2/3, 1/4)$. $X3$ is at $(0, 0, 1/4)$.
}
\begin{tabular}{lcc}
\hline
\textbf{ }	& \textbf{\ce{Pt3In2}}	& \textbf{\ce{Pt3Tl2}}\\
\hline
radiation	&	x-ray	&	neutron	\\
$a$ ({\AA})	&	5.5763(2)	&	5.63736(4)	\\
$c$ ({\AA})	&	13.6919(5)	&	13.8450(2)	\\
$x_{\rm{Pt}}$	&	0.169(1)	&	0.1667(8)	\\
$z_{\rm{Pt}}$	&	0.0890(2)	&	0.0854(1)	\\
$z_{X1}$	&	0.9132(6)	&	0.9055(2)	\\
Rp	&	10.8	&	10.5	\\
Rwp	&	15.7	&	6.71	\\
$\chi^2$	&	2.44	&	7.19	\\
\hline
\end{tabular}
\end{table}

This structure is similar to, but distinct from, the structure of the recently investigated double kagome net compounds \ce{Fe3Sn2} and \ce{V3Sb2}. The comparison is facilitated by examining Figures \ref{fig:structure}e and \ref{fig:structure}f, in which only the transition metal atoms are shown. The different stacking of the double kagome nets is apparent, and nicely illustrates the differences in the compounds' space group symmetries. In Pt$_3$\textit{X}$_2$ the $6_3$ screw axis rotates each double layer around the c-axis by 180$^{\circ}$ as they are stacked. In the Fe and V compounds, the stacking is defined by the rhombohedral centering, which simply translates each successive layer in the ab-plane relative to the previous layer.

To obtain reliable structural information and investigate the possibility of site disorder in Pt$_3$\textit{X}$_2$, Rietveld refinement of powder diffraction data was performed. Since Pt and In have significantly different atomic numbers, x-ray diffraction is suitable for this. Pt and Tl, however, have little x-ray scattering contrast, so neutron powder diffraction was used for this compound. The results are shown in Figure \ref{fig:diffraction} and structural parameters and agreement indices are summarized in Table \ref{table}.

The \ce{Pt3In2} sample contained a secondary phase identified to be \ce{Pt13In9}. The refinement indicated the concentration of this impurity to be near 3\% by weight. The peak shapes indicated the presence of anisotropic strain broadening, likely induced by grinding the sample. This was modeled in FullProf using the quartic form of the general strain formulation in Laue class $6/mmm$ and dramatically improved the fit. The \ce{Pt3Tl2} sample appeared single phase by neutron diffraction.

The structural parameters are in reasonable agreement with those previously reported \cite{bhan1968mischungen}. For both compounds, the refined x coordinate of the Pt atom is within experimental error of 1/6, which corresponds to an ideal kagome net. No evidence for significant site mixing was seen in the data. To test this, refinements were performed with In or Tl mixed onto the Pt site, and separately with Pt mixed onto the three In or Tl sites. For \ce{Pt3In2}, in the former case the composition of the Pt site refined to include only 6\% In. In the latter case, refinement returned a maximum amount of Pt on any In site of less than 4\%. Neither scenario resulted in an improvement in the fit quality judged by the $\chi^2$ values. For \ce{Pt3Tl2}, all the sites refined to their nominal stoichiometric occupancies within the estimated standard deviations, and $\chi^2$ was not improved by the site mixings. In addition, the overall average compositions of the two materials were determined by semiquantitative EDS measurements to be Pt$_{3.0(1)}$In$_{2.0(1)}$ and Pt$_{3.0(1)}$Tl$_{2.0(1)}$. This was calculated using spectra from 14 points across multiple \ce{Pt3Tl2} crystals and single points on 7 different grains of polycrystalline \ce{Pt3In2}. Thus, based on the present study, both \ce{Pt3In2} and \ce{Pt3Tl2} appear to be stoichiometric compounds with limited intrinsic disorder.

\begin{figure*}
\begin{center}
\includegraphics{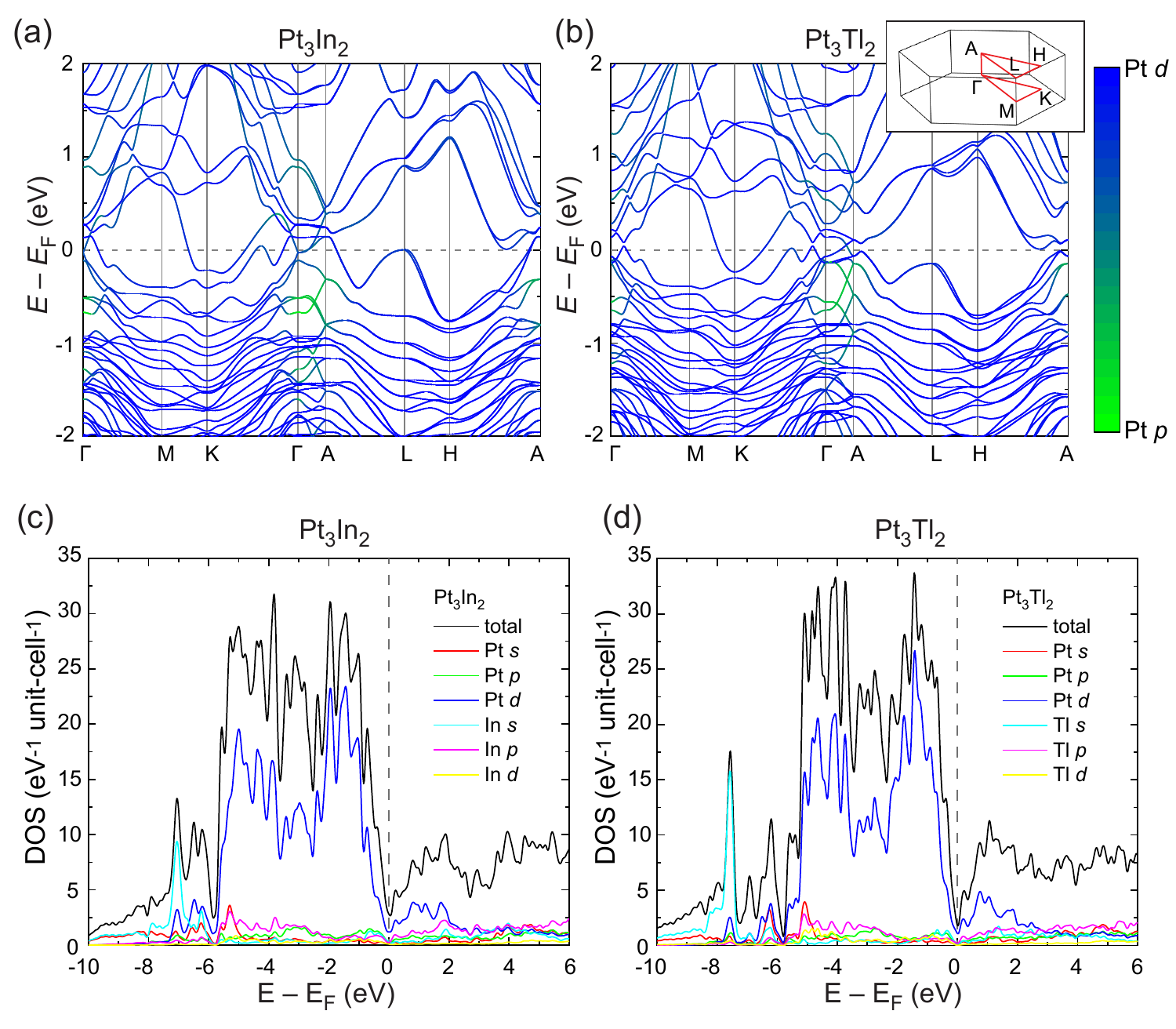}
\caption{\label{fig:DFT}
Electronic structure of Pt$_3$\textit{X}$_2$ from density functional theory calculations. Band structures are shown in (a) and (b), and the bands are colored based on their relative Pd-$d$ and Pt-$p$ character. Total and orbitally projected densities of states (DOS) are shown in (c) and (d), and are plotted on a per unit cell basis. One unit cell contains 12 Pt atoms and 8 In or Tl atoms (4 formula units). Dashed lines indicate the Fermi levels. The Brillouin zone and k-path used in the band structure diagrams are shown in the inset of (b).
}
\end{center}
\end{figure*}

The calculated electronic structures of \ce{Pt3In2} and \ce{Pt3Tl2} are shown in Figure \ref{fig:DFT}. As expected, states within a few eV of the Fermi level are dominated by Pt-$5d$ orbitals, especially the valence bands. In both compounds, multiple bands cross the Fermi level, with several hole-like pockets along the \textbf{k}-path traced in Figure \ref{fig:DFT}a and \ref{fig:DFT}b, and an electron-like pocket near the K point. The Fermi levels lie at pseudogaps in the densities of states (DOS). At $E_F$ the total DOS per unit cell (12 Pt aoms, 8 In or Tl atoms) is calculated to be 2.8 eV$^{-1}$ for \ce{Pt3In2} and 1.8 eV$^{-1}$ for \ce{Pt3Tl2}.

The flat bands, van Hove singularities, and Dirac nodes characteristic of kagome nets in the tight-binding and nearest-neighbor-hopping limits are not easily identifiable in Pt$_3$\textit{X}$_2$. This is likely due to the extended nature of the Pt-$5d$ orbitals, which results in strong covalent and beyond nearest neighbor interactions. This was similarly seen in CoSn-based compounds with heavier elements \cite{meier2020flat}, where the these features were much less distinct in PtTl when compared to CoSn and NiIn, or even to the $4d$ transition metal compound RhPb.

\begin{figure}
\begin{center}
\includegraphics[width=3.25in]{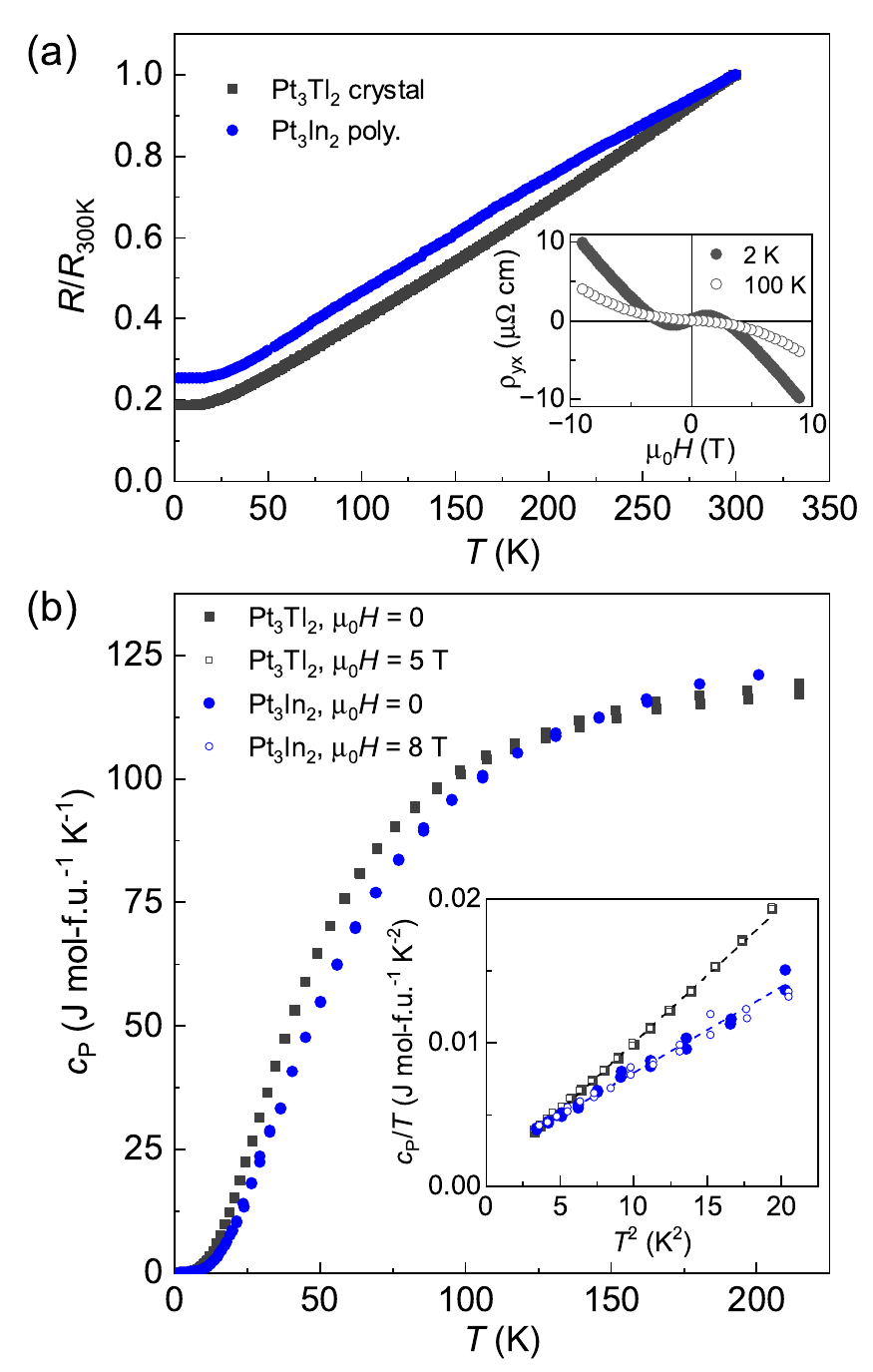}
\caption{\label{fig:rhocp}
Electrical transport and heat capacity results from Pt$_3$\textit{X}$_2$. The resistivities in (a) were measured in zero applied magnetic field and are shown normalized to their values at 300\,K. Room temperature values are $1.2 \times 10^{-4}~\Omega$cm for \ce{Pt3Tl2} (current in the plane) and on the order of $10^{-4}~\Omega$cm for \ce{Pt3In2} (polycrystalline). Transverse magnetoresistance (Hall effect) results for \ce{Pt3Tl2} are shown in the inset of (a). The heat capacity (b) was measured at the fields indicated in the legend using a single crystal of \ce{Pt3Tl2} and polycrystalline \ce{Pt3In2}. The low temperature heat capacity is shown in the inset of (b), with linear fits to $c_P/T$ vs $T^2$ used to extract the Sommerfeld coefficients and Debye temperatures.
}
\end{center}
\end{figure}

Results of electrical resistivity measurements are shown in Figure \ref{fig:rhocp}a. The data are plotted relative to values at 300\,K. The \ce{Pt3In2} sample was a piece cut from a polycrystalline pellet that had been pressed and sintered as described in the Materials and Methods section. The sample was small and not ideally shaped for accurately determining the cross sectional area for the current flow. This along with the polycrystalline nature of the sample make any precise value of the resistivity unreliable, but measurements near room temperature suggest the powder-averaged resistivity of \ce{Pt3In2} is in the $10^{-4}~\Omega$cm range. A \ce{Pt3Tl2} crystal was used for the measurements, and the current was applied in the ab-plane. The value measured at 300\,K was $1.2 \times 10^{-4}~\Omega$cm. Both compounds behave like metals with resistivity increasing approximately linearly with temperature above about 50\,K. The residual resistivity ratio defined as $R$(300\,K)/$R$(2\,K) is about 4 for \ce{Pt3In2} and about 5 for \ce{Pt3Tl2}.

The Hall effect was measured for \ce{Pt3Tl2} and the results are shown in the inset of Figure \ref{fig:rhocp}a. For these measurements the field $H$ was applied out of the plane of the crystal. The data are plotted as $\rho_{yx} = V_y t / I_x$, where the current $I$ and measured voltage $V$ are in the plane and $t$ is the thickness of the crystal. The observed nonlinearity of $\rho_{yx}$ vs H indicates the contribution from both electron and hole bands to the conduction. The Hall effect changes strongly with temperature as well, with the low field Hall coefficient changing sign between 100 and 2\,K.

The Hall effect data suggest a relatively complex Fermi surface, with the contributions from multiple carrier pockets changing with temperature. The resistivity data are consistent with typical behaviors seen in ``bad'' metals, semi-metals, or degenerately doped semiconductors. Both of these observations are consistent with the calculated electronic structures discussed above, which show the Fermi level resides in a pseudo-gap region but is crossed by multiple bands.

The heat capacity is shown in Figure \ref{fig:rhocp}b. The data show no indications of phase transitions, consistent with the transport results discussed above and the magnetization data below. At high temperatures the specific heat capacities ($c_P$) per mole of formula unit (f.u.) approach the Dulong-Petit limit of 3R per mole of atoms or 124.7\,J~mol-f.u.$^{-1}$K$^{-2}$. The low temperature behavior is shown in the inset of Figure \ref{fig:rhocp}b, with the data plotted as $c_P/T$ vs $T^2$. The linear fits shown on the plot give the Sommerfeld coefficient $\gamma$ from the intercept and the Debye temperature $\theta_D$ can be determined from the slope. For \ce{Pt3In2} this gives $\gamma$ = 2.0(2)\,mJ~mol-f.u.$^{-1}$K$^{-2}$ and $\theta_D$ = 254(6)\,K. For the \ce{Pt3Tl2} crystal this gives $\gamma$ = 0.52(5)\,mJ~mol-f.u.$^{-1}$K$^{-2}$ and $\theta_D$ = 218(1)\,K. The polycrystalline sample of \ce{Pt3Tl2} used for neutron diffraction was also measured, and gave similar values of $\gamma$ = 0.5(2)\,mJ~mol-f.u.$^{-1}$K$^{-2}$ and $\theta_D$ = 213(4)\,K. The Debye temperatures are relatively low, consistent with the heavy element constituents, and as expected somewhat higher in the In compound than the Tl compound. Note that the inset in Figure \ref{fig:rhocp}b contains data collected in zero field (used for the linear fitting) and also in relatively high applied magnetic fields. The results show no significant magnetic field effect on the heat capacity.

Comparing the values of $\gamma$ between the two compounds indicates a higher electronic density of states at the Fermi level in \ce{Pt3In2} than in \ce{Pt3Tl2}. This is supported by the electronic structure calculations in Figure \ref{fig:DFT}, and may partially explain why the resistivity values in the two samples are similar despite the polycrystalline nature of the \ce{Pt3In2} sample. The Sommerfeld coefficient can be calculated from the DOS at $E_F$ [$D(E_F)$] by $\gamma = (\pi^2/3)k_B^2D(E_F)$. Using $D(E_F)$ = 0.70 eV$^{-1}$f.u.$^{-1}$ and 0.45 eV$^{-1}$f.u.$^{-1}$ for the In and Tl compounds, respectively, gives $\gamma$(\ce{Pt3In2})\,=\,1.7\,mJ\,K$^2$f.u.$^{-1}$ and $\gamma$(\ce{Pt3Tl2})\,=\,1.1\,mJ\,K$^2$f.u.$^{-1}$. These values are similar to those extracted from the low temperature heat capacity data, and the similarities suggest the compounds have relatively weak electron-electron correlations.

\begin{figure}
\begin{center}
\includegraphics[width=3.25in]{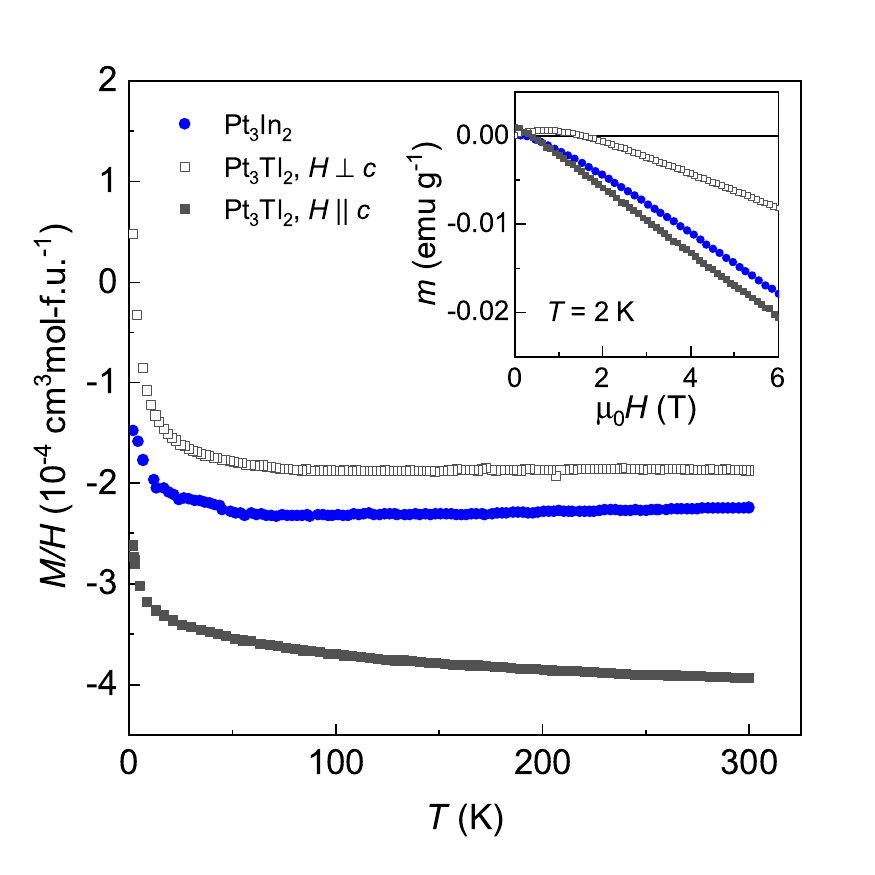}
\caption{\label{fig:mag}
Results of magnetization measurements on Pt$_3$\textit{X}$_2$. Magnetic susceptibility ($M/H$) was measured upon cooling in an applied field of $\mu_0H$\,=\,2\,T for a powder sample of \ce{Pt3In2} and 1\,T for single crystals of \ce{Pt3Tl2}. Isothermal magnetization curves are shown in the inset.
}
\end{center}
\end{figure}

The magnetic behaviors are summarized in Figure \ref{fig:mag}. The main panel shows the magnetic susceptibility of single crystals of \ce{Pt3Tl2} measured with the field applied parallel and perpendicular to the crystallographic c-axis and of polycrystalline \ce{Pt3In2}. The data have been corrected to account for contributions from the sample holders. All of the data sets contain Curie tails, low temperature upturns upon cooling, that may be attributed to low concentrations of local moments associated with defects or impurities. Otherwise, the data are negative and relatively temperature independent (some weak but notable $T$ dependence is seen for $H$ along the c-axis in \ce{Pt3Tl2}).

Estimating the core or Larmor diamagnetic contribution using the values tabulated for the noble gas cores in Ref. \citenum{AandM} gives $-1.9\times 10^{-4}$\,cm$^{3}$mol-f.u.$^{-1}$ for \ce{Pt3In2} and $-2.2\times 10^{-4}$\,cm$^{3}$mol-f.u.$^{-1}$ for \ce{Pt3Tl2}. The remaining magnetism is attributed to the conduction electrons. This includes Landau diamagnetism due to electronic orbits and Pauli paramagnetism due to electron spin.  Both are proportional to the density of states at the Fermi level. For a band with effective mass $m*$, this magnetic susceptibility is given by $\chi_0 = \mu_B^2D(E_F)[1-\frac{1}{3}(m_e/m*)^2]$, where the first term is the Pauli contribution and the second the Landau contribution. Since the Landau diamagnetism depends on the effective mass, it can overcome the positive Pauli contribution if the effective mass is small (less than $m_e/\sqrt{3}$ for a spherical Fermi surface) \cite{BlundellMagnetism}. It can also be anisotropic, and may account for the anisotropic behavior seen for \ce{Pt3Tl2} in Figure \ref{fig:mag} and in some related CoSn-type kagome compounds \cite{meier2020flat}.

The expected temperature independent susceptibility from the conduction electrons can be estimated using $D(E_F)$ values determined from the DFT calculations and assuming an effective mass of unity so that $\chi_0 = \frac{2}{3}\mu_B^2D(E_F)$. In this case the combined Pauli and Landau susceptibility $\chi_0$ (in units of cm$^3$mol-f.u.$^{-1}$) is calculated from $D(E_F)$ in units of eV$^{-1}$f.u.$^{-1}$ by $\chi_0 = 2.15\times10^{-5} D(E_F)$. This gives 1.5$\times10^{-5}$ cm$^3$mol-f.u.$^{-1}$ for \ce{Pt3In2} and 9.7$\times10^{-6}$ cm$^3$mol-f.u.$^{-1}$ for \ce{Pt3Tl2}.

The above estimates of the Larmor susceptibility calculated from the noble gas cores and the Pauli + Landau contributions calculated from $D(E_F)$ can be combined for comparison with the measured values in Figure \ref{fig:mag}. The calculated total susceptibility for \ce{Pt3In2} is $-1.8\times10^{-4}$ cm$^3$mol-f.u.$^{-1}$, and the measured value at room temperature is $-2.2\times10^{-4}$ cm$^3$mol-f.u.$^{-1}$. The calculated value for \ce{Pt3Tl2} is $-2.1\times10^{-4}$ cm$^3$mol-f.u.$^{-1}$. This should be compared to the isotropic average of the measured susceptibility given by $[2\chi(H \perp c) + \chi(H\|c)]/3$, which is $-2.6\times10^{-4}$ cm$^3$mol-f.u.$^{-1}$ at room temperature. Considering the approximations and assumptions, the calculated and experimental values compare quite favorably.

\begin{figure}
\begin{center}
\includegraphics[width=3.25in]{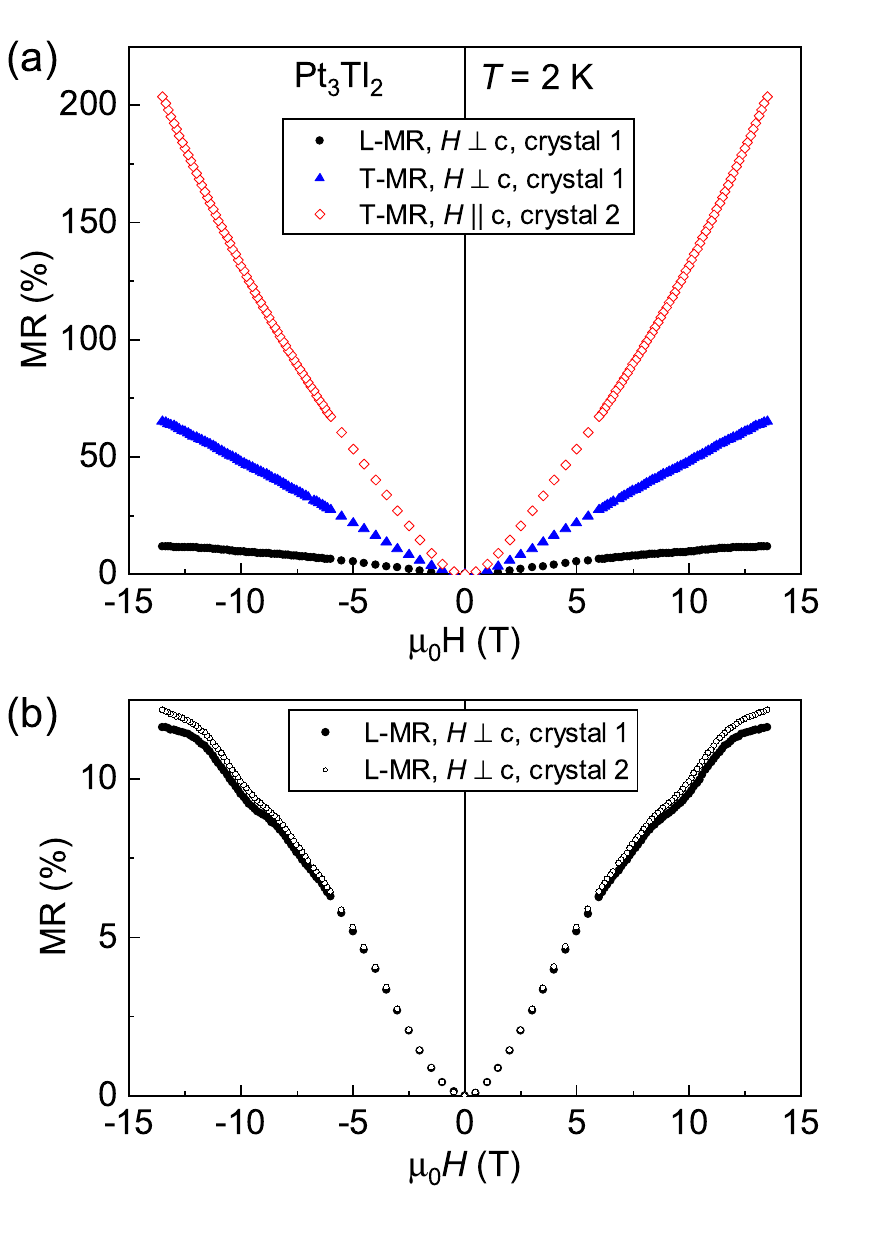}
\caption{\label{fig:MR}
Isothermal magnetoresistance data from \ce{Pt3Tl2} crystals. The magnetoresistance MR is defined as MR$(H) = [R(H) - R(0)]/R(0)$. The current was applied within the ab-plane ($\perp$c). For the longitudinal (L-MR) configuration the field was applied along the direction of the current. For the transverse configurations the field was applied along the c-direction (T-MR, $H||c$) and in the plane but perpendicular to the current direction (T-MR, $H\perp c$). Data for all three configurations, measured at $T$\,=\,2\,K, are shown in (a). L-MR data is shown in (b) from measurements on two different crystals.
}
\end{center}
\end{figure}

Since single crystals of \ce{Pt3Tl2} were available, anistropic magnetoresistance measurements were performed at 2\,K and fields up to $\pm$13.5\,T. For these measurements, the current $I$ was always applied in the plane of the platelike crystals ($I \perp c$). The field was applied in three directions relative to the current and the crystal. The longitudinal magnetoresistance (L-MR) was measured by applying the field parallel to the current ($H \| I$). The transverse magnetoresistance (T-MR) was measured by applying the field perpendicular to the current ($H \perp I$), with the field directed either out of the plane ($H \| c$) or in the plane ($H \perp c$). The measurements were performed at 2\,K and the results are summarized in Figure \ref{fig:MR}.

Each magnetoresistance curve shows interesting behaviors. The T-MR for $H \| c$ reaches a remarkably high value of 204\% with no indication of saturation. The T-MR with $H \perp c$ is approximately linear for fields above a few Tesla, and also reaches a high value of 65\%. Large magnetoresistance is often associated with high carrier mobility and/or semimetallike band structures with nearly compensated electron and hole pockets \cite{ali2014large}. The latter may apply to \ce{Pt3Tl2} based on the band structure and properties described above.

\begin{figure}
\begin{center}
\includegraphics[width=3.25in]{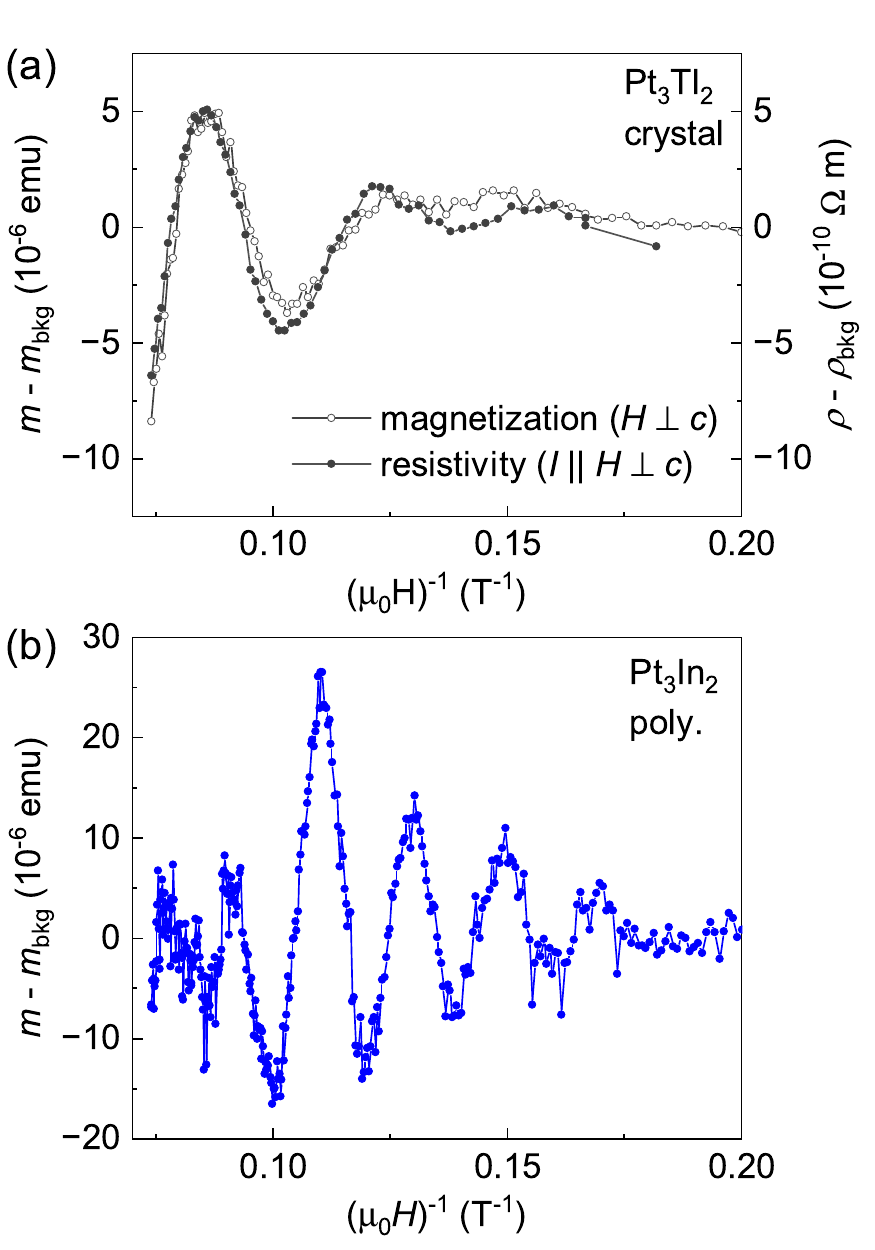}
\caption{\label{fig:osc}
Quantum oscillations in Pt$_3$\textit{X}$_2$ measured at 2\,K. (a) Oscillations of magnetization $m$ and resistivity $\rho$ in \ce{Pt3Tl2} crystals. The field was applied in the ab plane, and parallel to the current direction in the resistivity measurements. (b) Oscillations of the magnetization in a polycrystalline sample of \ce{Pt3In2}. Quadratic backgrounds ($\rho_{\rm{bkg}}, m_{\rm{bkg}}$) were subtracted from the measured data to isolate the oscillatory features. Note the data are plotted against 1/($\mu_0$H).
}
\end{center}
\end{figure}

The most remarkable features in the data are the undulations seen in the L-MR at high fields. This is shown in Figure \ref{fig:MR}b, where data measured on two different crystals are plotted. These are interpreted as quantum oscillations arising from the Shubnikov–de Haas effect, where the frequency of the oscillations (in 1/$\mu_0H$) relate to the area of closed extremal orbits on the Fermi surface \cite{Shoenberg1984}. These oscillations are not as easily observed in magnetoresistance data in other orientations, likely because of the overall larger field dependence of the resistivity in those cases.

A quadratic background was fit to the L-MR data from 4 to 13.5\,T and then subtracted to reveal the oscillatory nature. This is shown in Figure \ref{fig:osc}a, where the background subtracted data is plotted as a function of inverse magnetic field. The same physics that produces the Shubnikov–de Haas effect in transport data is expected to produce the de Haas-van Alphen effect in magnetization \cite{Shoenberg1984}. Therefore, the isothermal magnetization curve measured at 2\,K with $H \perp c$ was similarly examined. The resulting field dependence is plotted in Figure \ref{fig:osc}a and it matches well the oscillations in resistivity. Three local maxima can be seen in the resistance data, at 0.086, 0.122, and about 0.16\,T$^{-1}$. From this an approximate frequency of 27\,T can be estimated. The magnetization of \ce{Pt3In2} at 2\,K was also examined to see if quantum oscillations could be observed. The results are shown in Figure \ref{fig:osc}b. Oscillations are indeed seen, with a period of 0.020\,T$^{-1}$ or frequency of 50\,T.

\section{Summary and Conclusions}

Pt$_3$\textit{X}$_2$ ($X$\,=\,In and Tl) are double-layer kagome metals with structures similar to but distinct from that of the topologically nontrivial high temperature ferromagnet \ce{Fe3Sn2} and the density wave hosting compound \ce{V3Sb2}. Here we have reported results from our investigation of these materials, including the growth of \ce{Pt3Tl2} producing large single crystals and a thorough examination of the structures and electronic properties of the two compounds. Our main findings are (1) the materials are stoichiometric with limited site mixing between Pt and In/Tl, (2) no phase transitions are identified between room temperature and 2\,K, (3) the Fermi level lies in a pseudogap in the band structure of both compounds, (4) despite the absence of characteristic kagome-related band structure features interesting electronic behaviors are observed, including relatively large, linear, and non-saturating magnetoresistance and quantum oscillations related to the Shubnikov-de Haas and de Haas-van Alphen effects at 2\,K in the magnetoresistance and magnetization data, respectively. Because the electronic structures near the Fermi levels are dominated by Pt bands, these behaviors are necessarily related to the kagome networks in the compounds and expand our overall picture and understanding of kagome metal systems.

\section*{Acknowledgements}
This research was funded by the U.S. Department of Energy, Office of Science, Basic Energy Sciences, Materials Sciences and Engineering Division. This research at ORNL's Spallation Neutron Source was sponsored by the Scientific User Facilities Division, Office of Basic Energy Sciences, U.S. Department of Energy. This research used resources of the Compute and Data Environment for Science (CADES) at the Oak Ridge National Laboratory, which is supported by the Office of Science of the U.S. Department of Energy under Contract No. DE-AC05-00OR22725.


\end{document}